\documentclass[prd,aps,eqsecnum,amssymb,amsmath,nofootinbib,12pt]{revtex4}
\def\be{\begin{equation}}
\def\ee{\end{equation}}
\def\bea{\begin{eqnarray}}

\def\eea{\end{eqnarray}}

\begin{document}

\title{Radiation Reaction in Schwarzschild Spacetime: Retarded Green's Function
via Hadamard-WKB  Expansion}
\author{Paul R. Anderson$^1$\thanks{%
Email: anderson@wfu.edu} and B. L. Hu$^2$\thanks{%
Email: hub@physics.umd.edu} \\
$^1${\small {\it Department of Physics, Wake Forest University, Winston-Salem, NC  27109, USA}}\\
$^2${\small {\it Department of Physics, University of Maryland,
College Park, MD 20742, USA}}}

\begin{abstract}
An analytic method is given for deriving the part of the retarded Green's
function $v(x,x')$ that contributes to the tail term in the radiation
reaction force felt by a particle coupled to a massless minimally coupled
scalar field.  The method gives an expansion of $v(x,x')$ for small separations of the
points $x, x'$ valid for an
arbitrary static spherically symmetric spacetime. It is obtained
by using a WKB approximation for the Euclidean Green's function
for the massless minimally coupled scalar field and is
equivalent to the DeWitt-Schwinger expansion for $v(x,x')$.
The first few terms in this expansion are displayed
here for the case of Schwarzschild spacetime.
\end{abstract}

 \maketitle

\section{Introduction}
{\it This latest version has two new appendices A and B
containing respectively the two Errata in Ref.~\cite{a-h-e-1} and \cite{a-h-e-2} to the version
published in Physical Review D~\cite{a-h}. In this version we have made the corrections
indicated therein.}

The past several years have seen a resurgence of interest in the
gravitational radiation reaction problem in part because space based
gravitational wave detectors such as the Lasar Interferometer Space
Antenna (LISA) are expected to be able to detect the gravitational
radiation emitted when a compact object such as a stellar mass black
hole or neutron star spirals into a supermassive black
hole\cite{lisa1,lisa2}.  In such a situation the change in the
spacetime geometry of the supermassive black hole due to the
presence of the compact object and the gravitational radiation it
emits is negligible (except near the compact object) so it is
sufficient to compute the trajectory of the compact object in the
background geometry of the supermassive black hole\cite{lisa2}. The
radiation reaction under these simplifying conditions is known as
the ``self-force''.

The radiation reaction problem for a point charge radiating
electromagnetic waves in a curved space background was first
investigated by DeWitt and Brehme~\cite{dewitt-brehme} and later by
Hobbs~\cite{hobbs}. Gravitational radiation reaction for a moving
particle was considered by Mino et al \cite{minoetal} and Quinn and
Wald \cite{quinnwald}. The self-force on a particle interacting with
a massless minimally coupled scalar field was considered by Quinn
~\cite{quinn}. For a review see Ref.~\cite{poisson}. The scalar
field case is of interest primarily because it contains many of the
same features as the gravitational and electromagnetic cases while
being much easier to work with.  As found originally by DeWitt and
Brehme~\cite{dewitt-brehme} for the electromagnetic case, it is
always possible to break the radiation reaction force into two
parts, one that is local and which reduces to the standard
Abraham-Lorentz-Dirac expression in the flat space limit, the other
part which is nonlocal and consists of an integral of the retarded
Green's function over the past trajectory of the particle
~\cite{minoetal,quinnwald,quinn}. The nonlocal part is often called
the ``tail'' term.

Much effort has gone into the computation of the self-force in
Schwarzschild and Kerr spacetimes~\cite{dd,macgruder,vilenkin,smith-will,ori-1,ori-2,barack-ori,burko,lousto,barack,barack-burko,detweiler,nms,bl,pp,blu,bmnos,bo,mns,detweiler-whiting,detweiler-messaritaki-whiting}. In most schemes that do not
involve the weak field limit it must be computed numerically by
expanding either the retarded Green's function, or the field
itself, in terms of spherical harmonics in Schwarzschild or the
spin-weighted spheroidal harmonics in
Kerr~\cite{ori-1,ori-2,barack-ori,burko,lousto,barack,barack-burko,bl,blu,bmnos,bo,mns,detweiler-whiting,detweiler-messaritaki-whiting}.
However, a drawback of these methods is that it has not been
possible to separate out the local and nonlocal parts.  The
result is that there are divergences which must be regularized.
The usual way to treat these divergences is through subtractions
that occur at the level of the modes, resulting in a finite mode
sum. This is similar to adiabatic
regularization~\cite{parker,parker-fulling,fulling-parker,hu74,fulling-parker-hu}
introduced in the context of quantum field theory in curved
spacetime~\cite{dewitt,birrell-davies}.  Zeta function
regularization has also been used~\cite{lousto}

There is another way to obtain the tail term, at least in principle,
by computing the part of the retarded Green's function that
contributes to it. In this paper we propose to use the Hadamard
expansion of the retarded Green's function for this purpose. It is
known that this expansion is valid only when the points are close
together. The self force on the object at a particular spacetime
point has contributions from all points over its past trajectory. In
nonlocal processes it is not unreasonable to start by taking into
account contributions from points closest to the object as they
usually give a greater weight than those farther away. A quasilocal
expansion such as the one we are suggesting is the logical way to
start an analytic approximation. Even if it turns out not to capture
the dominant contribution, it is still worthwhile to investigate its
range of validity for the specific task at hand. If any such
analytic approximation produces even marginally reasonable results,
it can provide a relatively quick way of estimating the self-force
for a given trajectory.  If the results are accurate enough then it
can be used in place of brute force numerical integrations, thus
giving an immense economy of effort.

For a massless minimally coupled scalar field, the retarded Green's
function takes the Hadamard form ~\cite{ALN,Wald,phillips-hu,quinn}
\footnote{Our definitions for the Hadamard expansion are equivalent
to those of Ref.~\cite{ALN,Wald,phillips-hu}.  The conventions of
Ref.~\cite{quinn} differ from ours.  To obtain the convention for
$G_{\rm ret}$ in Ref.~\cite{quinn} let $G_{\rm ret} \rightarrow
G_{\rm ret}/(4 \pi)$.  Substituting this into Eq.\
(\ref{eq:hadamard}) and comparing with Eq.\ (9) of Ref.~\cite{quinn}
gives $u(x,x') = U(x,x')$, and $v(x,x') = 2 V(x,x')$.}
\begin{equation}
  G_{ret}(x,x') =  \theta(x,x') \left\{\frac{u(x,x')}{4 \pi } \delta[\sigma(x,x')] -
  \frac{v(x,x')}{8 \pi} \theta[-\sigma(x,x')] \right\}
\label{eq:hadamard}
\end{equation}
Here $\theta(x,x')$ is defined to be zero outside of the past light
cone and one inside of it.  The quantity $\sigma(x,x')$ is equal to
one-half the square of the proper distance between the points $x$
and $x'$ along the geodesic connecting them. The function $v(x,x')$
contributes to the tail part of the self-force. It obeys the
equation
\begin{equation}
\Box_x v(x,x') = 0 \;.
 \label{eq:boxv}
\end{equation}
Note that $v(x,x')$ is finite when the points come together.

For small separations of the points the functions $u$ and $v$ can
be computed by using the DeWitt-Schwinger
expansion~\cite{Schwinger,dewitt,christensen,christensen-thesis}.
To second order in derivatives of the metric the result for $v$ is
\begin{eqnarray}
v(x,x') &=& -\frac{1}{6} R(x) \;.
\label{eq:vds}
\end{eqnarray}
Christensen's expansions
\cite{christensen} can be used to
compute $v(x,x')$ to fourth order.  Phillips and Hu
\cite{phillips-hu} have computed it to sixth order for a
conformally coupled massless scalar field. Their result is useful
for the study of a massless minimally coupled field in
Schwarzschild spacetime because in any spacetime with zero scalar
curvature the Green's functions for the massless scalar field do
not depend upon the coupling to the scalar curvature. However,
taking  the general expressions for the terms in an arbitrary
spacetime from Ref.~\cite{phillips-hu} and evaluating them for the
Schwarzschild geometry is still  nontrivial.

The results of these calculations show that for Schwarzschild
spacetime the sixth order contribution to $v(x,x')$  from the
DeWitt-Schwinger expansion gives the first nonvanishing term, and
that is the only term which the results from Ref.~\cite{phillips-hu}
can provide. To compute the self-force more terms are needed from
an expansion for $v(x,x')$. Also, given the fact that no explicit
computations of $v(x,x')$ have been done for Schwarzschild
spacetime using the DeWitt-Schwinger expansion, it is of some
general interest to obtain the first few nonvanishing terms in
the series.

In this paper we describe a method we have developed to compute
an expansion for $v(x,x')$ for a massless minimally coupled scalar
field in a general static spherically symmetric spacetime.  The
method involves the use of a WKB expansion for the radial mode
functions of the Euclidean Green's function for the scalar field.
It gives the same results as the DeWitt-Schwinger expansion.
Because of its reliance on a WKB expansion for the radial modes
of the Euclidean Green's function, it should be straight-forward to
generalize the method to the cases of the electromagnetic and gravitational
fields in static spherically symmetric spacetimes.  Since the
scalar, electromagnetic, and gravitational wave equations
in Kerr spacetime are separable~\cite{carter,brilletal,teukolsky}, it is likely that the method can be
adapted to those cases as well.

In Sec. \ref{sec:euclidean} we show the relationship between the
Euclidean Green's function and $v(x,x')$ for the massless
minimally coupled scalar field in a general static spherically
symmetric spacetime. In Section \ref{sec:vcomp} our method of
deriving an expansion for $v(x,x')$ in powers of $(x-x')$ is given and
the first few terms of the expansion for
Schwarzschild spacetime are displayed.
We draw our conclusions in Section \ref{sec:concl}.

\section{Euclidean Green's Function and the Tail Term}
\label{sec:euclidean}

The metric for a static spherically symmetric spacetime can be written
in the form
\begin{equation}
ds^2 = -f(r) dt^2 + h(r) dr^2 + r^2 d\Omega^2 \;.
\label{eq:metric}
\end{equation}
where $f(r)$ and $h(r)$ are arbitrary functions of the radial
coordinate and $d\Omega^2$ is the metric of a 2-sphere. Changing
the time variable $t$ to $\tau = i t$ gives the Euclidean metric.
(Quantities defined in the Euclidean space carry a subscript $E$).
The Euclidean Green's function is a solution to the equation
\begin{equation}
\Box_x G_E(x,x') = - \frac{\delta (x,x')}{\sqrt{g_E}}
\label{eq:BoxGE}
\end{equation}
where $g_E$ is the determinant of the Euclidean metric. If the
analytic continuation
\begin{equation}
(\tau-\tau')^2 \rightarrow -(t-t')^2 + i \epsilon
\label{eq:analytic}
\end{equation}
is used with $\epsilon$ an infinitesimal positive real quantity then~\cite{gac}
\begin{equation}
G_E(-i\tau,\vec{x};-i\tau',\vec{x'}) = i G_F(x,x')
\label{eq:GEGF}
\end{equation}
with $G_F$ the Feynman Green's function. Since~\cite{christensen}
\begin{equation}
Im\, G_F(x,x') = -\frac{1}{2}\, G^{(1)}(x,x')
\end{equation}
where $G^{(1)}(x,x')$ is the Hadamard Green function, it is clear
that
\begin{equation}
Re\, G_E(-i\tau,\vec{x};-i\tau',\vec{x'}) = \frac{1}{2}
G^{(1)}(x,x') = \frac{u(x,x')}{8 \pi^2 \sigma(x,x')} +
  \frac{v(x,x')}{16 \pi^2} \log [|\sigma(x,x')|] + \frac{w(x,x')}{16 \pi^2}
  \label{eq:HadamardGE}
 \end{equation}
where in the second equality we have expressed $G^{(1)}(x,x')$ in
the Hadamard form~\cite{ALN,Wald,phillips-hu}.

Note that if the points are separated in the time direction, or if
they are farther apart in the time direction than in other
directions then~\cite{christensen-thesis,gac}
\begin{equation}
\sigma(x,x') = -\frac{1}{2} f(r) (t-t')^2 + O[(x-x')^3] \;.
\label{eq:sigma}
\end{equation}
Thus $v(x,x')$ is proportional to the coefficient of the
$\log[(\tau-\tau')^2]$ part of $G_E(x,x')$.  It is this fact that
will enable us to derive an expansion for $v$ by finding one for
$G_E$ using a WKB approximation.

To find such an expansion for $G_E$ first note that in a spacetime with metric
(\ref{eq:metric}) an exact expression for $G_E$ is~\cite{ahs}
\begin{equation}
  G_E(x,x')= \frac{1}{4 \pi^2} \,\int_0^\infty d \omega \cos[\omega(\tau-\tau')] \,
                  \sum_{\ell=0}^\infty (2 \ell + 1) P_\ell(\cos \gamma) C_{\omega \ell}\,
                  p_{\omega \ell}(r_<)\, q_{\omega \ell}(r_>) \;.
  \label{eq:GE}
\end{equation}
Here $P_\ell$ is the Legendre Polynomial of the $\ell$th order and
 $\cos \gamma \equiv \hat {\bf x} \cdot \hat {\bf x}'$ is the
direction cosine between the spatial vectors defining the points
$x, x'$. The notation $r_>$ ($r_<$) refers to the larger (smaller) of $r$ and $r'$.
 The mode function $p_{\omega \ell}$ satisfies the appropriate boundary
condition at some small value of $r$  and $q_{\omega\ell}$ satisfies the appropriate
boundary condition at some large value of $r$.  They are both solutions to the radial equation
\begin{equation}
\frac{1}{h} \frac{d^2 S}{dr^2} + \left[\frac{2}{r h} + \frac{1}{2 f h} \frac{d f}{dr}
               - \frac{1}{2 h^2} \frac{d h}{d r} \right] \frac{d S}{d r}
    - \left[\frac{\omega^2}{f} + \frac{\ell (\ell+1)}{r^2} \right]  S  \;\;= 0 \;.
\label{eq:mode}
\end{equation}
 They satisfy the Wronskian condition
\begin{equation}
C_{\omega \ell} \left[p_{\omega \ell} \frac{d q_{\omega \ell}}{d r} - q_{\omega \ell}
                  \frac{d p_{\omega \ell}}{d r} \right] = - \frac{1}{r^2} \left(\frac{h}{f} \right)^{1/2} \;.
\label{eq:wronskian}
\end{equation}

A WKB approximation for the modes can be derived as follows:
First define
\begin{eqnarray}
p_{\omega \ell} &=&  \frac{1}{(2 r^2 W)^{1/2}}\, \exp\left[\int^r W \left(\frac{h}{f}\right)^{1/2} d r \right] \;,  \nonumber \\
q_{\omega \ell} &=& \frac{1}{(2 r^2 W)^{1/2}}\, \exp\left\{-\left[\int^r W \left(\frac{h}{f}\right)^{1/2} d r \right]\right\}
\label{eq:wkbdef}
\end{eqnarray}
with
\begin{eqnarray}
W^2 &=& \Omega^2 + \frac{1}{2 r h} \frac{d f}{d r} - \frac{f}{2 r h^2} \frac{d h}{d r} \nonumber \\
  &+& \frac{1}{2} \left[\frac{f}{h W} \frac{d^2 W}{dr^2} + \left(\frac{1}{h} \frac{d f}{dr}
            - \frac{f}{h^2} \frac{d h}{d r} \right) \frac{1}{2 W} \frac{d W}{d r}
            - \frac{3}{2} \frac{f}{h} \left(\frac{1}{W} \frac{d W}{d r} \right)^2 \right]
\label{eq:W}
\end{eqnarray}
and
\begin{eqnarray}
\Omega^2 &=& \omega^2 + \ell (\ell + 1) \frac{f}{r^2}.
\label{eq:Omega}
\end{eqnarray}
Then solve the equation for $W$ iteratively, the lowest order
solution being $W = \Omega$.  Each iteration adds two derivatives
of the metric.  To second order one finds
\begin{eqnarray}
W &=& \Omega + \frac{1}{2 \Omega} \left(\frac{1}{2 r h} \frac{d f}{d r} - \frac{f}{2 r h^2} \frac{d h}{d r}\right)
  - \frac{1}{8 \Omega^3} \left(\frac{1}{2 r h} \frac{df}{dr} -
  \frac{f}{2 r h^2} \frac{dh}{dr} \right)^2
\nonumber \\
& & \;\; + \frac{1}{4} \left[\frac{f}{h \Omega^2} \frac{d^2 \Omega}{d r^2} +
      \left(\frac{1}{h} \frac{d f}{d r} - \frac{f}{h^2} \frac{d h}{d r} \right) \frac{1}{2 \Omega^2} \frac{d \Omega}{d r} - \frac{3}{2} \frac{f}{h} \frac{1}{\Omega^3} \left(\frac{d \Omega}{d r}\right)^2 \right]
\end{eqnarray}
Substitution of Eq.\ (\ref{eq:wkbdef}) into Eq.\ (\ref{eq:wronskian}) shows that
for the WKB ansatz, $C_{\omega \ell} = 1$.

\section{Computation of $v(x,x')$}
\label{sec:vcomp}

To compute an expansion for $v(x,x')$ in powers of $(x-x')$ one
first solves Eq.(\ref{eq:W}) to a specified adiabatic order and
substitutes the result into Eq. (\ref{eq:GE}).  {\bf As shown below,} the resulting
expression can be broken into sums and integrals of the form
\begin{equation}
S_{mnp} = \int_0^\infty d\omega \cos\omega(\tau-\tau')] \omega^{2n} \sum_{\ell=0}^\infty (2\ell+1) \frac{[l(l+1)]^m}{\Omega^p}
\label{eq:Sijk}
\end{equation}
with $m$, $n$, and $p$ integers.

Our goal is to determine the coefficient of the $\log(\tau-\tau')$
term since this is related to $v(x,x')$ through Eqs.\ (\ref{eq:HadamardGE})
and (\ref{eq:sigma}).  This can be accomplished
by using the Plana sum formula~\cite{plana}
\begin{equation}
\sum_{n=N}^\infty F(n) = \frac{1}{2} F(N) + \int_N^\infty dn F(n)
+ i \int \frac{dt}{e^{2\pi t}-1} \,[F(N+it) - F(N-it)]
\label{eq:plana}
\end{equation}
to compute the sum over $\ell$.  There are superficial divergences
in some of the sums over $\ell$ which can be isolated by expanding
the summand for a given sum in inverse powers of $\ell$ and then
truncating at order $1/\ell$. If the remaining terms are subtracted
from the summand then the result will be finite.  Effectively this
amounts to subtracting terms which are proportional to
$\delta(\tau-\tau')$ and its derivatives.  This procedure is
discussed in Ref.~\cite{ahs}. After computing the sum over $\ell$
one can expand the result in inverse powers of $\omega$. Most of the
expressions will be infrared divergent, however this is not
important because we are only looking for the coefficient of the
$\log(\tau-\tau')$ term.\footnote{There is an alternative way to do
the WKB expansion which gives no infrared
divergences~\cite{candelas-howard,howard,ahs}.} Thus it suffices to
put in a lower limit cutoff $\lambda$ on the integral over $\omega$.

The integral over $\omega$ can be computed for each term of the
series.  For positive powers of $\omega$ there are no
$\log(\tau-\tau')$ terms so these do not contribute to $v(x,x')$.
For all negative powers of $\omega$ there will be $\log(\tau-\tau')$ terms.
For $\omega^{-1}$ one finds
\begin{eqnarray}
\int_\lambda^\infty d\omega \cos[\omega(\tau-\tau')]
\frac{1}{\omega} &=& -{\rm ci}(\lambda (\tau-\tau')) =
     -\log(\tau-\tau') + ... \;\;.
\label{eq:ci}
\end{eqnarray}
For all negative powers of $\omega$, successive integrations by
parts can be performed until the integral is in the form
(\ref{eq:ci}).  The result is
\begin{equation}
\int d\omega \cos[\omega(\tau-\tau')] \frac{1}{\omega^{2n+1}} =
   \frac{(-1)^{n+1}}{(2n)!}\,(\tau-\tau')^{2n} \log(\tau-\tau') + ... \;\;.
\end{equation}
If the points are split only in the time direction then one finds that using
a second order WKB expansion gives $v$ to zeroth order in $(t-t')$,
a fourth order one gives $v$ to second order in $(t-t')$, and so forth.  For
Schwarzschild spacetime the lowest nonvanishing order is $(t-t')^4$ which
requires use of a sixth order WKB approximation.

To compute the expansion when the points are also split in the
angular direction one can begin by expanding the angular part of
the mode functions in powers of $(\cos\,\gamma - 1)$ so that
\begin{equation}
P_\ell(\cos\,\gamma) = 1 + \frac{\ell(\ell+1)}{2} (\cos\,\gamma - 1) + ...
\end{equation}
Then the sum over $\ell$ and the integral over $\omega$ can be
computed as before.  At the end of the calculation  one should
also expand the terms of the form $(\cos\,\gamma -1)^j$ in powers
of $\theta-\theta'$ and $\phi-\phi'$.

To compute the expansion when the points are also split in the
radial direction one should first fix $r'$ to be either less than
or greater than $r$.  For our purpose it will not matter which is
chosen so, as an example, we assume that $r' < r$.  Then one
expands $p_{\omega \ell}(r')$ in powers of $r'-r$ and repeatedly
uses the mode equation (\ref{eq:mode}) to eliminate all but first
derivatives of $p_{\omega \ell}$.  Thus

\[ p_{\omega \ell}(r') = p_{\omega \ell}(r) + p_{\omega
\ell}'(r) (r'-r) \]
\begin{equation} + \left\{\left[-\frac{2}{r} - \frac{1}{2 f} \frac{d f}{dr}
               + \frac{1}{2 h} \frac{d h}{d r} \right] p_{\omega \ell}'(r)
    + \left[\frac{\omega^2 h}{f} + \frac{\ell (\ell+1) h}{r^2} \right] p_{\omega \ell}(r)
       \right\} \frac{(r'-r)^2}{2} + ... .
\end{equation}
Then this expansion is substituted into the expression
(\ref{eq:GE}) for $G_E$ and the WKB expansion is introduced to
write this equation in terms of $W$ and its derivatives as before.
The terms will be proportional to various powers of $\omega$ and
$\ell(\ell+1)$ multiplying either $ 1, 1/W$ , or $\;
W'/W^2 \;$.  Those multiplying $1$ are proportional to
$\delta(\tau-\tau') \delta(\Omega-\Omega')$ and various
derivatives of these delta functions.  Therefore, they do not
contribute to the calculation of $v(x,x')$ and can be ignored.
For the other terms one substitutes the WKB
approximation for $W$ to some specified order  and computes the
coefficients of the $\log(\tau-\tau')$ terms as before.

To affirm the validity of the WKB scheme, we note that when this
program is carried out for a Reissner-Nordstr\"{o}m spacetime the
expansion for $v(x,x')$ agrees with the results of
Christensen~\cite{christensen} at order $(x-x')^2$.  To leading order, it is also a
solution to Eq.\ (\ref{eq:boxv}).  For Schwarzschild
spacetime $v(x,x')$ vanishes at order $(x-x')^2$. However, by using an
eighth order WKB approximation for the modes it has been possible
to compute $v(x,x')$ to order $(x-x')^6$ in Schwarzschild spacetime.
Writing
\begin{equation}
  v(x,x') = \sum_{i,j,k = 0}^\infty v_{ijk} (t-t')^{2i} \, (\cos \gamma - 1)^j
  \, (r-r')^k
\end{equation}
we find
\begin{equation}
 v_{000} = v_{001} = v_{100} = v_{010} = v_{002} = v_{101} =
 v_{011} = v_{003} = 0
 \end{equation}
 and
\begin{eqnarray}
v_{200} &=& \frac{3\,M^2\,{\left( 2\,M - r \right)
}^3}{224\,r^{11}}
\nonumber \\
v_{110} &=& \frac{27\,M^2\,{\left( 2\,M - r \right) }^2}{280\,r^8}
\nonumber \\
v_{102} &=& \frac{-9\,M^2\,\left( 2\,M - r \right) }{560\,r^9}
\nonumber \\
v_{020} &=& \frac{9\,M^2\,\left( 2\,M - r \right) }{280\,r^5}
\nonumber \\
v_{012} &=& \frac{-9\,M^2}{280\,r^6}
\nonumber \\
v_{004} &=& \frac{3\,M^2}{1120\,\left( 2\,M - r \right) \,r^7}
\nonumber \\
v_{201} &=&
  \frac{3\,M^2\,\left( 11\,M - 4\,r \right) \,{\left( 2\,M - r \right) }^2}
   {224\,r^{12}}
\nonumber \\
v_{111} &=&  \frac{27\,M^2\,\left( 8\,M - 3\,r \right) \,\left( 2\,M
- r \right) }
   {280\,r^9}
\nonumber \\
v_{103} &=& \frac{-9\,M^2\,\left( 9\,M - 4\,r \right)
}{560\,r^{10}}
\nonumber \\
v_{021} &=& \frac{9\,M^2\,\left( 5\,M - 2\,r \right) }{280\,r^6}
\nonumber \\
v_{013} &=& \frac{-27\,M^2}{280\,r^7}
\nonumber \\
v_{005} &=&
  \frac{3\,M^2\,\left( 7\,M - 4\,r \right) }
   {1120\,{\left( 2\,M - r \right) }^2\,r^8}
\nonumber \\
v_{300} &=&
  \frac{M^2\,{\left( 2\,M - r \right) }^3\,
     \left( 39\,M^2 - 26\,M\,r + 4\,r^2 \right) }{480\,r^{15}}
\nonumber \\
v_{210} &=&
  \frac{M^2\,{\left( 2\,M - r \right) }^2\,
     \left( 135\,M^2 - 91\,M\,r + 14\,r^2 \right) }{224\,r^{12}}
\nonumber \\
v_{202} &=&
  \frac{M^2\,\left( 2\,M - r \right) \,
     \left( 171\,M^2 - 136\,M\,r + 26\,r^2 \right) }{224\,r^{13}}
\nonumber \\
v_{120} &=&
  \frac{9\,M^2\,\left( 2\,M - r \right) \,
     \left( 21\,M^2 - 18\,M\,r + 4\,r^2 \right) }{560\,r^9}
\nonumber \\
v_{112} &=&
  \frac{3\,M^2\,\left( 567\,M^2 - 473\,M\,r + 94\,r^2 \right) }{560\,r^{10}}
\nonumber \\
v_{104} &=&
  \frac{-3\,M^2\,\left( 51\,M - 28\,r \right) \,\left( 5\,M - 2\,r \right) }
   {1120\,\left( 2\,M - r \right) \,r^{11}}
\nonumber \\
v_{030} &=&
  \frac{M^2\,\left( 3\,M - 14\,r \right) \,\left( 2\,M - r \right) }
   {840\,r^6}
\nonumber \\
v_{022} &=&
  \frac{3\,M^2\,\left( 81\,M^2 - 70\,M\,r + 14\,r^2 \right) }
   {560\,\left( 2\,M - r \right) \,r^7}
\nonumber \\
v_{014} &=&
  \frac{- M^2\,\left( 729\,M^2 - 773\,M\,r + 202\,r^2 \right) }
   {1120\,{\left( 2\,M - r \right) }^2\,r^8}
\nonumber \\
v_{006} &=&
  \frac{M^2\,\left( 249\,M^2 - 292\,M\,r + 86\,r^2 \right) }
   {3360\,{\left( 2\,M - r \right) }^3\,r^9}
\end{eqnarray}
This expression is a solution to Eq.\ (\ref{eq:boxv}) to
$O\left[(x-x')^4\right]$.  By that we mean that if it is substituted
into the left hand side of Eq.\ (\ref{eq:boxv}) then the resulting
expression will be zero to $O\left[(x-x')^4\right]$.

\section{Conclusions}
\label{sec:concl}

We have presented a method that allows for the calculation of the
part of the retarded Green's function $v(x,x')$ that contributes
to the tail part of the radiation reaction force for a massless
minimally coupled scalar field in a general static spherically
symmetric spacetime.  We expect it to be straight-forward to
adapt this method to the cases of
the electromagnetic and gravitational fields in static spherically
symmetric spacetimes.  It may be possible to adapt it, for all three fields,
to Kerr spacetime as well given that the wave equations for all three
fields are separable in the Kerr background~\cite{carter,brilletal,teukolsky}.

We have explicitly calculated $v(x,x')$ in Schwarzschild spacetime
to $O\left[(x-x')^6\right]$ for an arbitrary separation
of the points. One indication of the correctness of these
expressions is that $v(x,x')$ satisfies Eq.\ (\ref{eq:boxv}) to the
appropriate order.

Although we have not yet computed enough terms to get an estimate of
the self-force,
one qualitative feature in our results is distinct from other
related cases~\cite{poisson2}.
In the large $r$ limit one can see that the
leading order terms in $v$ are all proportional to $M^2$. By
comparison, calculations by DeWitt and DeWitt~\cite{dd} and
Pfenning and Poisson~\cite{pp} show that when the spacetime
curvature is everywhere small, such as is the case for a static
star, then the leading order term is proportional to $M$. More
specifically, in the calculation of DeWitt and DeWitt the metric
is everywhere Schwarzshild except at the origin where they assume
a delta function mass source. They made the approximation that the
Green's function can be well approximated everywhere by the flat
space Green's function (this is known to be false for regions
close to the black hole event horizon)and found that the leading
order term comes from a signal that propagates to the central
condensation at $r=0$, bounces off and comes back to the current
location of the particle. 

One might be concerned that similar processes will contribute a
leading order term that is linear in $M$ for the black hole case.
This may well turn out to be the case, and one should consider this
factor seriously.   However, before drawing any direct implications
we caution that the case of a particle orbiting a static star is
qualitatively different from that of a particle orbiting a black
hole.  For example,  the calculations in Ref.~\cite{dd,pp} assume
that the curvature is nowhere large which is not a valid assumption
for the case of a particle orbiting a black hole. Also,  in the case
of a black hole the existence of an event horizon precludes
classical waves from scattering off the central condensation (note
this is different from the superradiance effect occurring in the
ergosphere of a Kerr black hole), even though such scattering can
occur for orbits with $r>3M$ everywhere.   We think it likely that
this type of nonlocal (infrared) contribution to the self-force can
be, at least approximately, decoupled from the quasilocal
contribution we are considering here.  If this is the case, then the
nonlocal contribution can be considered separately  and the result
can be added to the quasilocal contribution to the self-force.

The method we have developed to compute $v(x,x')$ can be used to
find the expansion to higher orders in $(x-x')$. Work is in progress
to compute an expansion for $v(x,x')$ to substantially higher order
in Schwarzschild spacetime, which should provide a better test of
the usefulness of the Hadamard expansion. Although the quasilocal
expansion may prove insufficient to compute the tail part of the
self-force, the facts that no regularization is necessary and the
resulting expression is analytic make an investigation of its
usefulness well worth the effort.  Such an investigation is underway
and the results will be presented elsewhere.

\section*{Acknowledgments}
We would like to thank Nicholas Phillips for helpful conversations,
Payman Eftekharzadeh for detecting (non-propagating) errors in two
equations of our original draft, and Eric Poisson for pointing out
the likely existence of a linear M term in the expression for the
self-force, which the present approximation lacks. This work was
supported in part by the National Science Foundation under grant
numbers PHY-0070981 and PHY-0300710.


\newpage

\appendix

\section{First Erratum}

In the Physical Review D (PRD) version of this paper~\cite{a-h}
there were a few notational and transcription errors that have been
corrected in this arXiv version and also published in an
erratum~\cite{a-h-e-1}.  For completeness we list the corrections in
that erratum here as well.

\begin{enumerate}
\item In Eq.\ (1.1) of the PRD version the general expression for the Hadamard form of
the retarded Green's function was given incorrectly.  It should read
\begin{equation}
  G_{\rm ret}(x,x') =  \theta(x,x') \left\{\frac{u(x,x')}{4 \pi } \delta[\sigma(x,x')] -
  \frac{v(x,x')}{8 \pi} \theta[-\sigma(x,x')] \right\}
\label{eq:hadamard}
\end{equation}

\item The relationships between our definitions of $u(x,x')$ and
$v(x,x')$ above and $U(x,x')$ and $V(x,x')$ of Ref.~\cite{quinn}
were given incorrectly in the PRD version. To obtain the correct
relationships first note that the convention for $G_{ret}$ in
Ref.~\cite{a-h} differs from that of Ref.~\cite{quinn}.  To obtain
that of Ref.~\cite{quinn} let $G_{\rm ret} \rightarrow G_{\rm
ret}/(4 \pi)$. Substituting this into Eq.\ (\ref{eq:hadamard}) and
comparing with Eq.\ (9) of Ref.~\cite{quinn} gives $u(x,x') =
U(x,x')$, and $v(x,x') = 2 V(x,x')$.

\item In Eq.\ (3.9) of the PRD version the expressions for the
following quantities should be multiplied by a minus sign:
$v_{201}$, $v_{111}$, $v_{103}$, $v_{021}$, $v_{013}$, and
$v_{005}$.
\end{enumerate}

These errors do not affect any other equations or results in the
paper and the above corrections have been made in this arXiv
version.

  We would like to thank Ardeshir Eftekharzadeh (April 2006) for
first pointing out the errors in Eq.~(3.9) of the PRD version, and
Barry Wardell (April 2007) for independently pointing them out. P.
R. A. would like to thank Hebrew University and the Department of
Theoretical Physics at the University of Valencia for hospitality.
This work was supported in part by the National Science Foundation
under grant numbers PHY03-00710 and PHY05-56292. P.R.A. acknowledges
the Einstein Center at Hebrew University, the Forchheimer
Foundation, and the Spanish Ministerio de Educaci\'on y Ciencia for
financial support.

\newpage

\section{Second Erratum}

In the Physical Review D (PRD) version of this paper~\cite{a-h}
there was a factor of 2 error that has been
corrected in this arXiv version and also published in an
erratum~\cite{a-h-e-2}.  For completeness we list the corrections in
that erratum here as well.

In Ref.~\cite{a-h} there was an error in the calculation of the coefficients in the Hadamard-WKB expansion of
$v(x,x')$.  Each coefficient in Eq.\ (3.9) should be multiplied by a factor or $2$.  A transcriptional error
in certain coefficients was corrected in~\cite{a-h-e-1}.  The expansion for $v$ is
\begin{equation}
  v(x,x') = \sum_{i,j,k = 0}^\infty v_{ijk} (t-t')^{2i} \, (\cos \gamma - 1)^j
  \, (r-r')^k  \;.
\end{equation}
The corrected coefficients, including both the corrections in~\cite{a-h-e-1} and the factor of 2, are
\begin{equation}
 v_{000} = v_{001} = v_{100} = v_{010} = v_{002} = v_{101} =
 v_{011} = v_{003} = 0
 \end{equation}
 and
\begin{eqnarray}
v_{200} &=& \frac{3\,M^2\,{\left( 2\,M - r \right)
}^3}{224\,r^{11}}
\nonumber \\
v_{110} &=& \frac{27\,M^2\,{\left( 2\,M - r \right) }^2}{280\,r^8}
\nonumber \\
v_{102} &=& \frac{-9\,M^2\,\left( 2\,M - r \right) }{560\,r^9}
\nonumber \\
v_{020} &=& \frac{9\,M^2\,\left( 2\,M - r \right) }{280\,r^5}
\nonumber \\
v_{012} &=& \frac{-9\,M^2}{280\,r^6}
\nonumber \\
v_{004} &=& \frac{3\,M^2}{1120\,\left( 2\,M - r \right) \,r^7}
\nonumber \\
v_{201} &=&
  \frac{3\,M^2\,\left( 11\,M - 4\,r \right) \,{\left( 2\,M - r \right) }^2}
   {224\,r^{12}}
\nonumber \\
v_{111} &=&  \frac{27\,M^2\,\left( 8\,M - 3\,r \right) \,\left( 2\,M
- r \right) }
   {280\,r^9}
\nonumber \\
v_{103} &=& \frac{-9\,M^2\,\left( 9\,M - 4\,r \right)
}{560\,r^{10}}
\nonumber \\
v_{021} &=& \frac{9\,M^2\,\left( 5\,M - 2\,r \right) }{280\,r^6}
\nonumber \\
v_{013} &=& \frac{-27\,M^2}{280\,r^7}
\nonumber \\
v_{005} &=&
  \frac{3\,M^2\,\left( 7\,M - 4\,r \right) }
   {1120\,{\left( 2\,M - r \right) }^2\,r^8}
\nonumber \\
v_{300} &=&
  \frac{M^2\,{\left( 2\,M - r \right) }^3\,
     \left( 39\,M^2 - 26\,M\,r + 4\,r^2 \right) }{480\,r^{15}}
\nonumber \\
v_{210} &=&
  \frac{M^2\,{\left( 2\,M - r \right) }^2\,
     \left( 135\,M^2 - 91\,M\,r + 14\,r^2 \right) }{224\,r^{12}}
\nonumber \\
v_{202} &=&
  \frac{M^2\,\left( 2\,M - r \right) \,
     \left( 171\,M^2 - 136\,M\,r + 26\,r^2 \right) }{224\,r^{13}}
\nonumber \\
v_{120} &=&
  \frac{9\,M^2\,\left( 2\,M - r \right) \,
     \left( 21\,M^2 - 18\,M\,r + 4\,r^2 \right) }{560\,r^9}
\nonumber \\
v_{112} &=&
  \frac{3\,M^2\,\left( 567\,M^2 - 473\,M\,r + 94\,r^2 \right) }{560\,r^{10}}
\nonumber \\
v_{104} &=&
  \frac{-3\,M^2\,\left( 51\,M - 28\,r \right) \,\left( 5\,M - 2\,r \right) }
   {1120\,\left( 2\,M - r \right) \,r^{11}}
\nonumber \\
v_{030} &=&
  \frac{M^2\,\left( 3\,M - 14\,r \right) \,\left( 2\,M - r \right) }
   {840\,r^6}
\nonumber \\
v_{022} &=&
  \frac{3\,M^2\,\left( 81\,M^2 - 70\,M\,r + 14\,r^2 \right) }
   {560\,\left( 2\,M - r \right) \,r^7}
\nonumber \\
v_{014} &=&
  \frac{- M^2\,\left( 729\,M^2 - 773\,M\,r + 202\,r^2 \right) }
   {1120\,{\left( 2\,M - r \right) }^2\,r^8}
\nonumber \\
v_{006} &=&
  \frac{M^2\,\left( 249\,M^2 - 292\,M\,r + 86\,r^2 \right) }
   {3360\,{\left( 2\,M - r \right) }^3\,r^9}
\end{eqnarray}

\acknowledgments   We would like to thank Barry Wardell for alerting us to the
possibility that a factor of two error existed in our results and for help
in finding it.  We would also like thank  Ardeshir Eftekharzadeh for help
in finding the error.  P. R. A. would like to thank the Physics Department at
the University of Bologna for hospitality.  This work was
supported in part by the National Science Foundation under grant
numbers PHY03-00710 and PHY05-56292.

\end{document}